\documentclass{ws-procs975x65}

\begin{document}

\markboth{Kov\'acs, Gergely, Horv\'ath}{Canonical analysis of radiating
atmospheres...}

\title{CANONICAL ANALYSIS OF RADIATING ATMOSPHERES OF STARS IN EQUILIBRIUM 
\footnote{
Research supported by OTKA grants no. T046939 and TS044665, the J\'{a}nos
Bolyai Fellowships of the Hungarian Academy of Sciences, the Pierre Auger
grant 05 CU 5PD1/2 via DESY/BMF and the EU Erasmus Collaboration between the
University of Szeged and the University of Bonn. Z.K. and L.\'A.G. 
thank the organizers of the 11th Marcel Grossmann Meeting for support.} 
}

\author{ZOLT\'AN KOV{\'{A}}CS$^{\dagger \ddagger}$, 
L\'ASZL\'O \'A. GERGELY$^{\ddagger}$ and 
ZSOLT HORV\'{A}TH$^{\ddagger}$}

\address{$\dagger$ Max-Planck-Institut f{\"u}r Radioastronomie,\\
Auf dem H\"ugel 69, D-53121 Bonn, Germany\\
$\ddagger$ Departments of Theoretical and Experimental Physics, University of Szeged,\\
D\'om t\'er 9, H-6720 Szeged, Hungary\\
\email{zkovacs@mpifr-bonn.mpg.de, gergely@physx.u-szeged.hu, 
zshorvath@titan.physx.u-szeged.hu}}

\begin{abstract}
The spherically symmetric, static spacetime generated by a cross-flow of
non-interacting null dust streams can be conveniently interpreted as the
radiation atmosphere of a star, which also receives exterior radiation.
Formally, such a superposition of sources is equivalent to an anisotropic
fluid. Therefore, there is a preferred time function in the system, defined
by this reference fluid. This internal time is employed as a canonical
coordinate, in order to linearize the Hamiltonian constraint. This turns to
be helpful in the canonical quantization of the geometry of the radiating
atmosphere.
\end{abstract}


\bodymatter

\section*{}
The quantum theory of gravitational collapse motivated many authors to study
models with both in- and outgoing thin null dust shells in a spherically 
symmetric geometry. Such models can equally apply for other phenomena, like 
radiative domains around stars in thermodynamical equilibrium. 
The model of a radiative stellar atmosphere composed of two null dust streams 
provides good prospects for carrying out a complete canonical analysis and  
quantization. We present here an overview of the Hamiltonian description of 
two cross-streaming radiation fields with spherical symmetry 
and the first steps towards the Dirac quantization of this
constrained Hamiltonian system.

Letelier demonstrated that the energy-momentum tensor of two superimposed,
counter-propagating radiation fields is equivalent to the 
energy-momentum tensor of a specific anisotropic fluid.\cite{Letelier} 
Based on this algebraic equivalence we have recently shown that the 
dynamics derived by extremizing the matter Lagrangians of
these two models are the same.\cite{HKG} 
For the purpose of canonical analysis the two
cross-flowing radiation fields can therefore be substituted with a single
anisotropic fluid (with radial pressure equaling the energy density and no 
tangential pressures).

The equivalence with the fluid model is crucial for our purposes since earlier
works on the Hamiltonian formalism of two cross-flowing radiation fields
with spherical symmetric geometry, although achieving important results,
could not solve the problem of the absence of an internal time. 
\cite{BicakHajicek} The possibility of replacing the two-component null dust
with an anisotropic fluid raises the possibility to introduce the proper time 
as an internal time in the Hamiltonian formalism, in analogy with the case of 
the incoherent dust.\cite{BrownKuchar} 

We foliate the static and spherically symmetric geometry by the spherically 
symmetric leaves $\Sigma _{t}$ labelled by the parameter time $t$: 
\begin{equation}
ds^{2}=-(N-\Lambda N^{r2})dt^{2}-\Lambda ^{2}N^{r}2dtdr+\Lambda
^{2}dr^{2}+R^{2}d\Omega ^{2}\;,  \label{ds20}
\end{equation}%
where $\Lambda (t,r)$ and $R(t,r)$ are the metric functions and $N$ and $%
N^{r}$ are the lapse function and the non-vanishing component of the shift
vector, respectively \cite{BCMN}. 

A static, spherical symmetric space-time
describing the cross-flow of two null dust streams (or equivalently an
anisotropic fluid) has been found \cite{Gergely}:%
\begin{equation}
ds^{2}=-2ae^{Z^{2}}R^{-1}(Z)[dT^{2}-R^{2}(Z)dZ^{2}]+R^{2}(Z)d\Omega ^{2}\;,
\label{ds21}
\end{equation}%
where $T$ and $Z$ are time and radial coordinates of the fluid particles and%
\[
-R(Z)=a\left[ e^{Z^{2}}-2Z\left( B+\int^{L}e^{x^{2}}dx\right) \right] \;.
\]%
Motivated by this exact solution we chose the scalar fields $\Lambda $, $R$, 
$T$ and $Z$ appearing in the metrics (\ref{ds20}) and (\ref{ds21}) as the
canonical coordinates of the gravity and the matter source. The proper time $%
T$ of the fluid particles provides the internal time for the colliding
radiation fields, whereas the radial coordinate $Z$ gives the Lagrangian
coordinate of the fluid particles for constant $\theta $ and $\phi $.

In order to provide the Hamiltonian description of this model, we perform
the Legendre transformation of the Lagrangian 
\[
S^{2ND}[{}^{(4)}g_{ab},\rho ]=\int d^{4}x\sqrt{{}^{(4)}g_{ab}}\rho
(u_{a}u^{a}+v_{a}v^{a})\;,
\]%
describing two non-interacting null dust streams with time-independent
energy density $\rho $, which propagate along the null congruences $u^{a}$
and $v^{a}$. We perform the transformation by decomposing the tangent
vectors of the two null congruences with respect to the gradients of the
matter variables, 
\begin{equation}
u_{a}=WT_{,a}+RWZ_{,a}\;,\qquad v_{a}=WT_{,a}-RWZ_{,a}\;,\qquad
W=ae^{Z^{2}}R\;,  \nonumber
\end{equation}%
and introducing the momenta $P$ and $P_{Z}$ canonically conjugated to $T$ and 
$Z$, 
\begin{equation}
P={\mathcal{G}}N^{-1}(T_{,t}-N^{r}T_{,r})\;,\qquad P_{Z}={\mathcal{G}}%
R^{2}N^{-1}(Z_{,t}-N^{r}Z_{,r})\;\qquad {\mathcal{G}}=2a\sqrt{g}\rho W^{2}. 
\nonumber
\end{equation}%
The matter Lagrangian can be then rewritten in the "already Hamiltonian"
from 
\begin{equation}
L^{2ND}=\dot{T}P+\dot{Z}P_{Z}-NH_{\bot }^{2ND}-N^{r}H_{r}^{2ND}~,  \nonumber
\end{equation}%
where the super-Hamiltonian and supermomentum constraints of the system
consisting of the two null dust streams are 
\begin{equation}
H_{\bot }^{2ND}=\mathcal{G}^{-1}(P^{2}+P_{Z}^{2}/R^{2})+\mathcal{G}%
(T^{\prime 2}+R^{2}Z^{\prime 2})\;,\qquad H^{2ND}=T^{\prime }P+Z^{\prime
}P_{Z}\;.  \nonumber
\end{equation}%
By eliminating the comoving density $\rho $ form the Hamiltonian constraint
and employing that the super-Hamiltonian and the supermomentum constraints
of the total system weakly vanish, 
\begin{equation}
H_{\bot }:=H_{\bot }^{G}+N_{\bot }^{2ND}\approx 0\;,\qquad
H_{r}:=H_{r}^{G}+N_{r}^{2ND}\approx 0\;,  \label{HbotHr}
\end{equation}%
we are able to solve the constraints with respect to the momenta $P_{T}$ and 
$P_{Z}$. The vacuum constraints $H_{\bot }^{G}$
and $H_{r}^{G}$ are expressed in terms of the preferred canonical variables
of spherically symmetric vacuum gravity,\cite{Kuchar} with $\Lambda $ and its canonical
momentum is replaced with the Schwarzschild mass $M$ and its canonical
momentum $P_{M}$. After solving the constraints with respect
to the momenta we can introduce a new set of linearized constraints,
equivalent to Eq. (\ref{HbotHr}), in which the momenta of the matter
variables are separated from the rest of the canonical data \cite{HKG}: 
\begin{equation}
H_{\uparrow }:=P+h[M,R,T,Z,P_{M},P_{R}]=0\;,\quad H_{\uparrow
Z}:=P_{Z}+h_{Z}[M,R,T,Z,P_{M},P_{R}]=0\;.  \nonumber
\end{equation}%
The above linearized form of the constraints is advantageous for two
reasons. First, the Hamiltonian constraint $H_{\uparrow }$ is resolved with
respect to the momentum $P$ canonically conjugated to the internal time $T$.
Second, the new constraints have strongly vanishing Poisson brackets and as
such form an Abelian algebra instead of the Dirac algebra of the old
constraints.

In the canonical quantization of gravity coupled to the two null dust
streams with spherically symmetric geometry the super-Hamiltonian constraint
becomes an operator equation on the state functional $\Psi \lbrack Z,T,M,R]$
of gravity, restricting the allowed states. Since classically the
super-Hamiltonian constraint was resolved with respect to the momentum $P$,
the operator condition leads to the functional Schr{\"{o}}dinger equation 
\begin{equation}
i\frac{\delta }{\delta T}\Psi \lbrack T,M,R]=h[M,R,T,Z,P_{M},P_{R}]\Psi
\lbrack T,M,R]\;.  \label{Sch}
\end{equation}%
The operator version of the supermomentum constraint $H_{\uparrow Z}$
applied on the state functional ensures that the quantum states are
independent of the dust frame $Z$ \cite{BrownKuchar}. Besides the Hilbert
space structure of the solutions to the Eq. (\ref{Sch}), the other advantage
of the linearized constraints is that their Abelian algebra turns into a
true Lie algebra of vacuum gravity. These promising achievements point
towards a possible consistent canonical quantization of the presented
superposed null dust system.

\bibliographystyle{ws-procs975x65}
\bibliography{Kovacs-GT7}

\end{document}